\pdfoutput=1
\documentclass[aps,prb,amsmath,amssymb,twocolumn]{revtex4-1}
\usepackage[utf8x]{inputenx}
\usepackage{microtype}
\usepackage{graphicx}
\usepackage{hyperref}

\begin{document}

\title{Probe-assisted spin manipulation in one-dimensional quantum dots}

\author{Yasha Gindikin and Vladimir A. Sablikov}

\affiliation{Kotel'nikov Institute of Radio Engineering and Electronics,
Russian Academy of Sciences, Fryazino, Moscow District, 141190,	
Russia}

\begin{abstract}
We study a spin structure that arises in a one-dimensional quantum dot with zero total spin under the action of a charged tip of a scanning probe microscope in the presence of a weak magnetic field. The evolution of the spin structure with changing the probe position is traced to show that the movable probe can be an effective tool to manipulate the spin. The spin structures are formed when the probe is located in certain regions along the dot due to the Coulomb interaction of electrons as they are redistributed between the two sections in which the quantum dot is divided by the potential barrier created by the probe. There are two main states: spin-polarized and non-polarized ones. The transition between them is accompanied by a spin precession governed by the Rashba spin-orbit interaction induced by the electric field of the probe. In the transition region the spin density changes strongly while the charge distribution remains nearly unchanged.
\end{abstract}

\maketitle

Development of the tools to operate and control the electron spin in quantum nanostructures is a challenging problem which attracts a good deal of attention nowadays\cite{PhysRevB.90.235311,PhysRevLett.110.086804,shulman2012demonstration,PhysRevB.86.085423,bluhm2011dephasing} and gives rise to a wide stream of physical researches of the spin-dependent phenomena in quantum dots (QD).\cite{doi:10.1146/annurev-conmatphys-030212-184248,hanson2007Spifewquadot,loss1998Quacomquadot} One of the most promising directions of searching for the effective mechanisms of the spin manipulation is based on the investigations of the spin dynamics in the presence of the spin-orbit interaction (SOI) and magnetic field. In low-dimensional quantum structures the electron-electron (\textit{e-e}) interaction also strongly affects the spin degree of freedom, but the effects caused by the combined action of the SOI and the \textit{e-e} interaction are little studied to date.

Effective manipulation of the spin is attained in a system of double QDs coupled by a potential barrier in the presence of a magnetic field by tuning the barrier potential.\cite{nadj2010spin,bertrand2014quantum} An important parameter of the spin-state switching is the energy difference between the spin states which determines the fidelity of their initialization and read-out. The manipulation mechanism in double QD system is associated with the singlet-triplet transitions (STTs).\cite{ashoori1993N-egrostaenequadotmagfie,vanderwiel1998Sintrafewquadot} For this mechanism the fidelity depends on the SOI strength, which determines the energy splitting of the singlet and triplet levels $\Delta_{SO}$. Since the SOI is usually not very strong in semiconductor quantum wires, this energy turns out to be not high.

In recent years a growing attention is paid to the methods using the tip of a scanning probe microscope (SPM) for studying the electron system in low-dimensional quantum structures and manipulating electrons.\cite{topinka2000imaging,woodside2002scanned,bleszynski-jayich2008ImaoneInAquadotInAnan,boyd2011scanning,ziani2012signatures,mantelli2012non,zhukov2014investigations,PhysRevB.91.075313} The manipulation of electrons can be very effective in one-dimensional (1D) QDs as the charged tip creates a negative potential that divides the QD into two quantum wells.\cite{qian2010} Under this condition the \textit{e-e} interaction strongly affects the electron density redistribution between the wells. Due to the \textit{e-e} interaction the tip moving along the 1D QD forces the electrons to pass one by one from one well to another. Qian \textit{et al.} studied this process ignoring the electron spin.\cite{qian2010}

We draw attention to the fact that the electrically charged probe makes it possible to manipulate the spin as well if an external magnetic field is present. An advantage of this mechanism of the spin manipulation is the tunability of the system parameters, which allows one to change the energy level difference between the wells and the inter-well tunnel coupling. The magnitude of the SOI can also be controlled because an essential contribution to SOI comes from the probe electric field that becomes rather high as the tip approaches the QD.

In the present paper we study the spin manipulation in a 1D QD containing an even number of electrons with zero total spin using the charged SPM tip. We consider the QD isolated from the reservoirs as this allows one to avoid the parasitic effects commonly accompanying the spin manipulation, such as the photon assisted tunneling.\cite{bertrand2014quantum} We trace the evolution of the electron charge and spin density distribution under the change in the probe position and potential.

The spin manipulation only becomes possible due to an interplay between the strong \textit{e-e} interaction and the high probe potential in the presence of an external magnetic field exceeding some critical value $B_c$. The probe potential divides the QD into two tunnel-coupled quantum wells. The Coulomb repulsion of electrons occupying the same quantum well produces strong Hubbard-like correlations due to which an electron once having occupied the narrowest well with spin parallel to $B$ blocks the electrons with opposite spin in the other well. In addition, the \textit{e-e} interaction leads to the formation of Wigner molecules\cite{hausler1993interacting,gindikin2007deformed} and shrinks the level spacing in the energy spectrum of the many-particle states with different spin configuration,\cite{hanson2007Spifewquadot} thereby significantly reducing the value of $B_c$.\cite{gindikin2011electron} The Rashba SOI induced by the probe electric field determines the spin dynamics in the course of switching between different spin states.

We solve this problem for the arbitrary \textit{e-e} interaction strength and probe position by using the exact diagonalization method. We find that the displacement of the probe along the QD results in the abrupt switching of the spin states of the QD between the non-polarized and polarized states. In the transition region between the polarized and non-polarized states the spin precession happens, while the spatial charge distribution remains essentially unchanged.

To be specific, consider a 1D QD containing four electrons in the presence of a charged probe. The QD extends in the $x$-direction, the wave function $\Psi$ obeys the open boundary conditions at the ends of the QD at $x=0$ and $x=L$. The magnetic field $B$ is directed along the $z$-axis. If the transverse quantization energy is larger that any energy scale under consideration then the Hamiltonian is
\begin{eqnarray}
H=\sum_{i=1}^{4}\left[\frac{p_{x_i}^2}{2m}+\frac{1}{2}g\mu_B B\sigma_{z_i}-e\left(\phi_{\mathrm{pr}}(x_i)+V(x_i)\right)\right]&\nonumber \\
+\sum_{i>j}U(x_i-x_j)+ H_\mathrm{SO}\;,&
\label{Hamiltonian}
\end{eqnarray}
where $p_{x_i}$ is the momentum of $i$-th electron, $\sigma_z$ is the Pauli matrix, $g$ is the effective $g$-factor. The Coulomb \textit{e-e} interaction potential can be approximated as
\begin{equation}
U(x_1-x_2)=\frac{U}{\varepsilon}\frac{e^2}{\sqrt{(x_1-x_2)^2+a^2}}\;,
\end{equation}
where $\varepsilon$ is the dielectric constant, $a$ is the QD transverse size, $U$ is an auxiliary dimensionless parameter that is introduced to study the effect of the \textit{e-e} interaction strength on the spin structure ($U=1$ unless stated otherwise). The electron-ion interaction potential $V(x)$ is determined by the pair potential $U(x_1-x_2)$ in the jelly model. The probe potential in the QD equals
\begin{equation}
  \phi_{\mathrm{pr}}(x)=\frac{Q}{\varepsilon}\left[(x-x_0)^2+z_0^2\right]^{-1/2},
\end{equation}
with $(x_0,0,z_0)$ being the probe position, Q -- the probe charge. The SOI Hamiltonian is taken as 
\begin{equation}
 H_\mathrm{SO}=\frac{\alpha}{\hbar} \sum_{i=1}^{4}\frac12 (E_{z}(x_i)p_{x_i}+p_{x_i}E_{z}(x_i))\sigma_{y_i},
\label{H_SOI}
\end{equation}
where $\alpha$ is the SOI parameter, $E_z(x)$ is the $z$-component of the probe electric field.

The ground state wave function $\Psi$ is found by exact diagonalization of the Hamiltonian~(\ref{Hamiltonian}). The spin density components, in the units of $\hbar/2$, are defined as $s_{\gamma}(x)=\langle \sum_{i=1}^4 \sigma_{\gamma_i}\delta(x-x_i) \rangle$, with $\gamma \in \{x,y,z\}$, average taken over the ground state. Due to open boundary conditions $s_y(x)=0$.

The system parameters are chosen to correspond to the InAs quantum dots, specifically, $\alpha=117$ $e\mbox{\AA{}}^2$, the electron effective mass is $m=0.0265\,m_0$, the static dielectric function is $\varepsilon=15$, the Bohr radius is $a_B=300$ \AA{}.\cite{martienssen2006springer} The bulk value of gyromagnetic factor $g=-15$ is assumed.\cite{zielke2014anisotropic} The system length is $L=900$ \AA{}, the probe charge is $Q=5e$, the height of the probe above the wire is $z_{\mathrm{probe}}=50$ \AA{}, the magnetic field $B=0.44$~T, which corresponds to the cyclotron frequency of $\hbar\omega_{c}=0.024 E_0$, with $E_0=1.72$ meV being the longitudinal quantization energy.

\begin{figure}
\centerline{\includegraphics[width=0.9\linewidth]{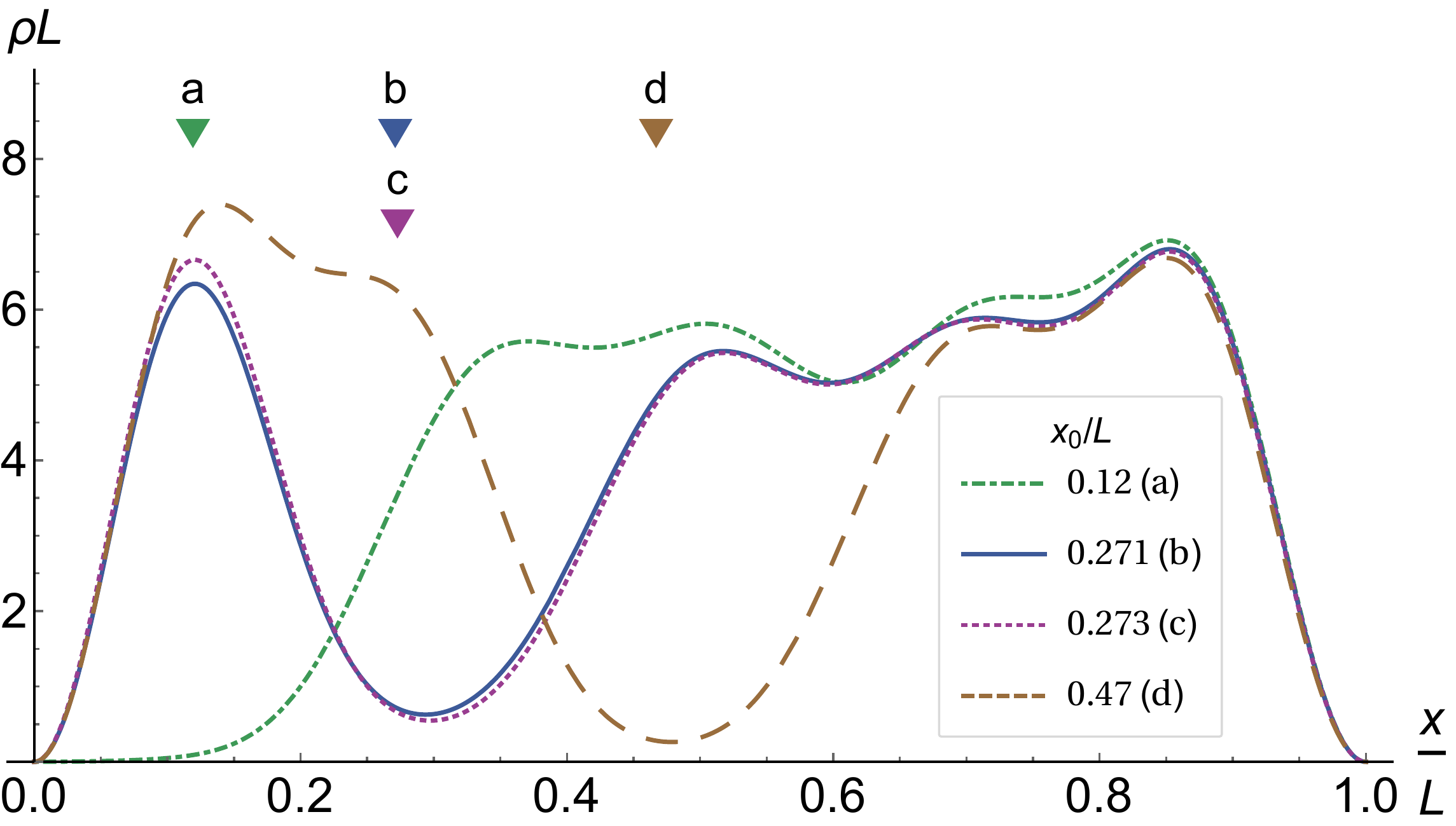}}
\caption{(Color online) The spatial dependence of the electron charge density in a 1D QD subjected to the action of a charged probe for several probe positions $x_0$, specified in the legend and depicted by the corresponding triangle.}
\label{fig1}
\end{figure}

\begin{figure}
\begin{tabular}{c}
\centerline{\includegraphics[width=0.85\linewidth]{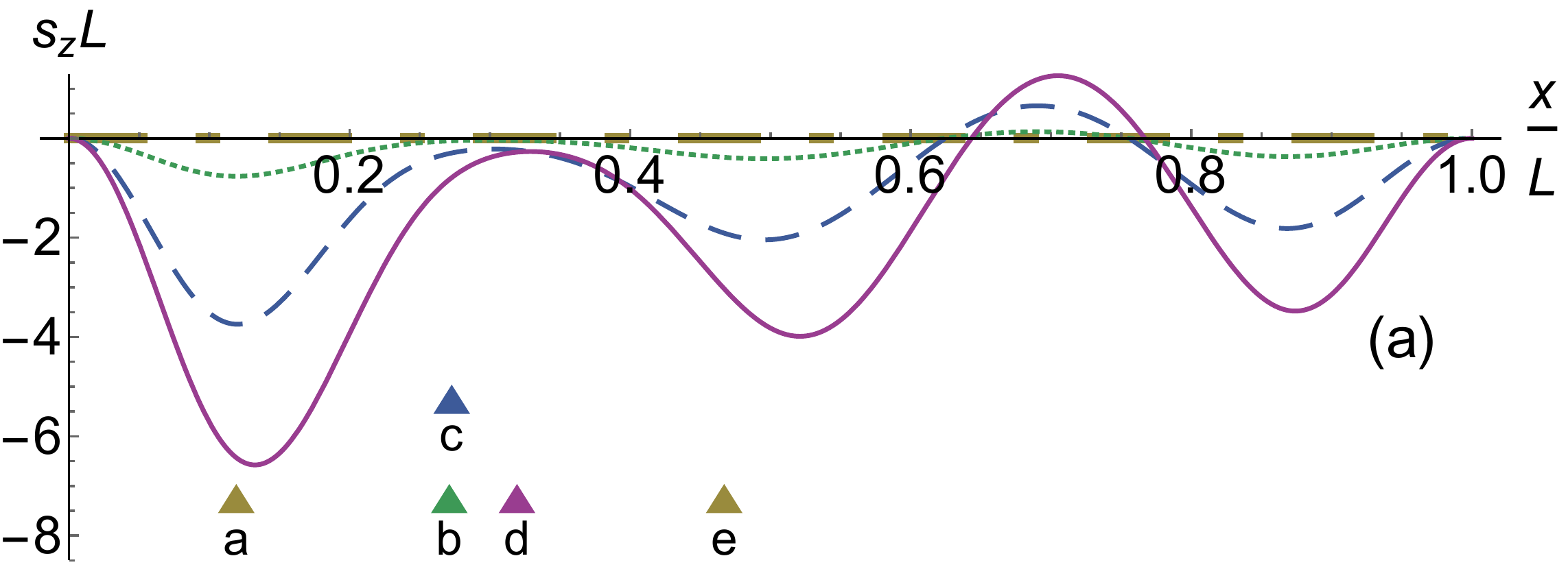}}\\
\centerline{\includegraphics[width=0.9\linewidth]{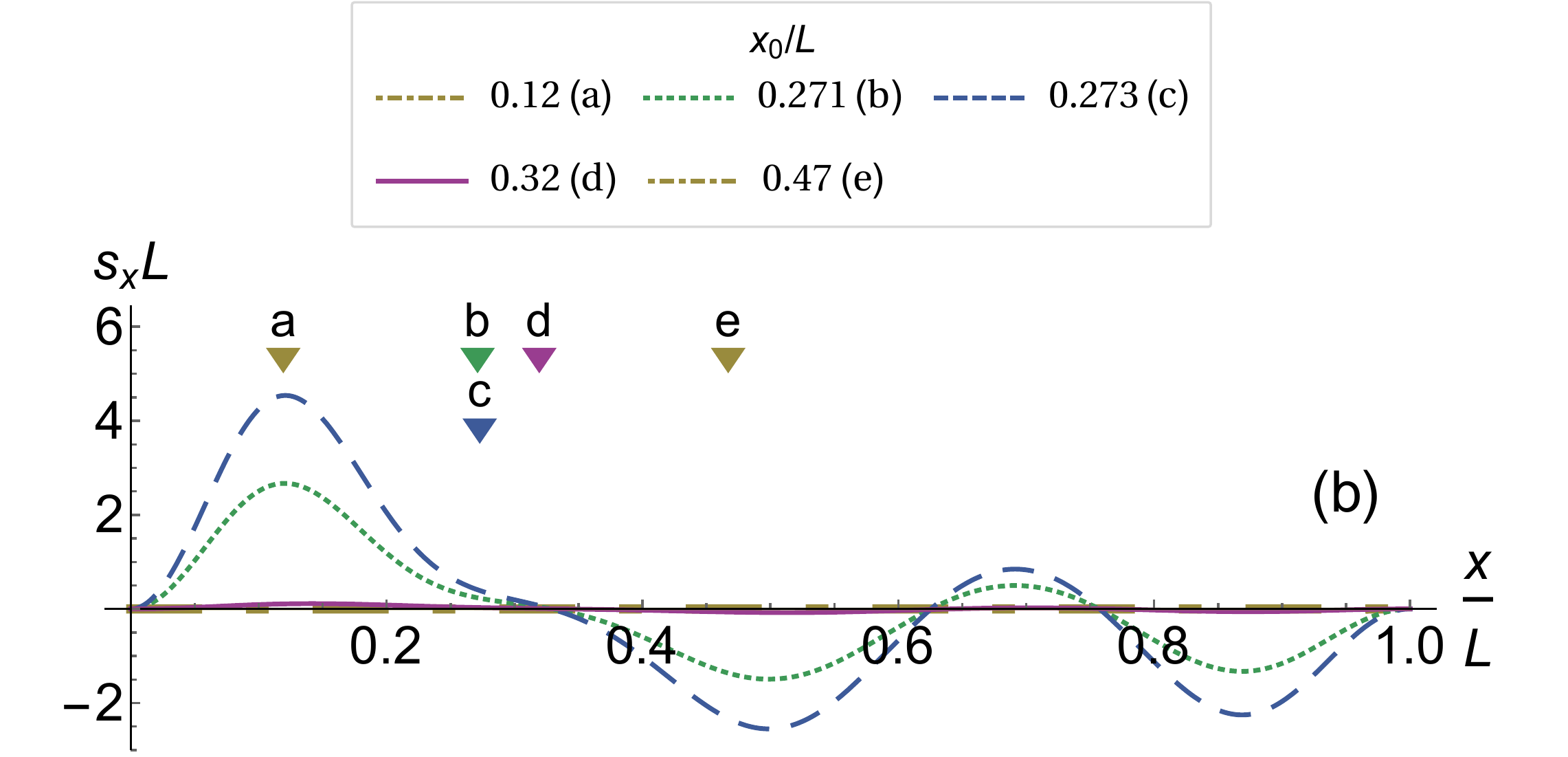}}
\end{tabular}
\caption{(Color online) The spatial dependence of the $z$-component (a) and $x$-component (b) of the spin density for several positions $x_0$ of the probe.}
\label{fig2}
\end{figure}

The electron density distribution along the QD is shown in Fig.~\ref{fig1} for four positions of the charged probe. When the probe is located at one (left) end of the QD, the electrons form a Wigner molecule. As the probe shifts from left to right, the electrons are squeezed to the right section of the QD, with spin density components being close to zero. At some critical probe position ($x_0\approx0.27 L$) first electron passes to the left side of the probe. This transition is followed by the emergence of the $z$-component of the spin density, reaching its maximum amplitude as the probe continues moving to the right as can be seen from Fig.~\ref{fig2}a. The switching of the spin state happens in a narrow range of the probe position, the width $\delta x_0$ of which is roughly estimated as $(mx_0^3/\hbar^2) \Delta_\mathrm{SO}$. We emphasize that the charge density $\rho(x)$ remains nearly constant in the transition region (see curves (b) and (c) in Figs.~\ref{fig1},~\ref{fig2}), which can be seen as a signature of the spin-charge separation.

The appearance of $s_z(x)$ at the critical probe position is accompanied by the emergence of the $x$-component of the spin density $s_x(x)$ as shown in Fig.~\ref{fig2}b. Note that $s_x(x)$ is non-zero only in the transition region of the width of $\delta x_0$. On the contrary, the $s_z(x)$ spin component exists within a band of a large enough width $\Delta x_0$, which depends on the magnitude of the \textit{e-e} interaction strength and magnetic field. The band corresponds to the state in which one electron with spin directed along the $z$-axis is localized in one of the quantum wells (a more narrow of the two). The band edges are determined by an energy balance under the variation of the probe position. The variation of the kinetic energy due to the electron localization in the quantum wells is compensated by the gain in the Coulomb interaction energy and the energy of the spin in the magnetic field. Three other electrons remain localized in the wider quantum well. They exhibit an antiferromagnetic ordering while there is a ferromagnetic correlation between the localized electron and the adjacent one (Fig.~\ref{fig2}a).

Upon the further increase of $x_0$ the second electron passes to the left of the probe, see the dashed line (d) in Fig.~\ref{fig1}. It carries the spin which is exactly opposite to the spin of the localized electron so that the spin in each of the quantum wells on both sides of the probe turns to zero in accordance with the Lieb-Mattis theorem. At the end of the first band where $s_z(x)\ne 0$, the $s_x(x)$ component appears again in the narrow region of the probe positions.

\begin{figure}
\begin{tabular}{c}
\centerline{\includegraphics[width=0.9\linewidth]{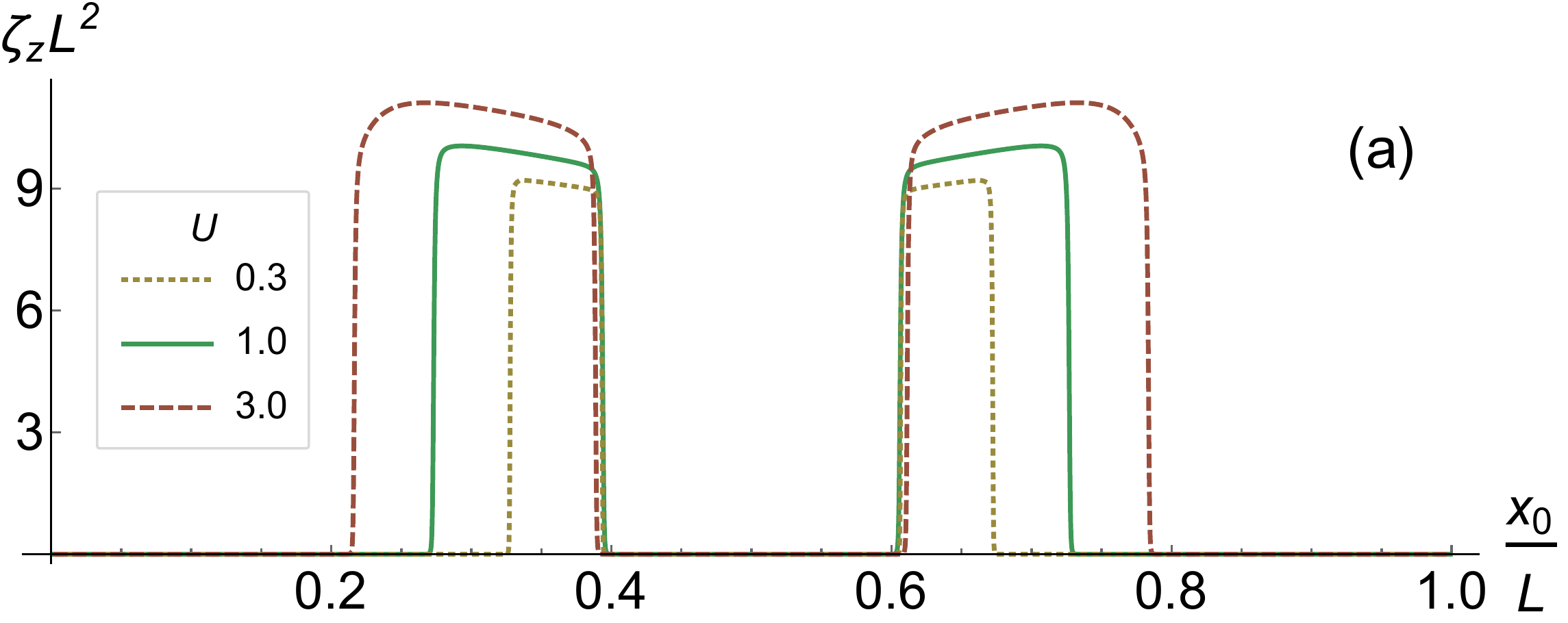}}\\
\centerline{\includegraphics[width=0.9\linewidth]{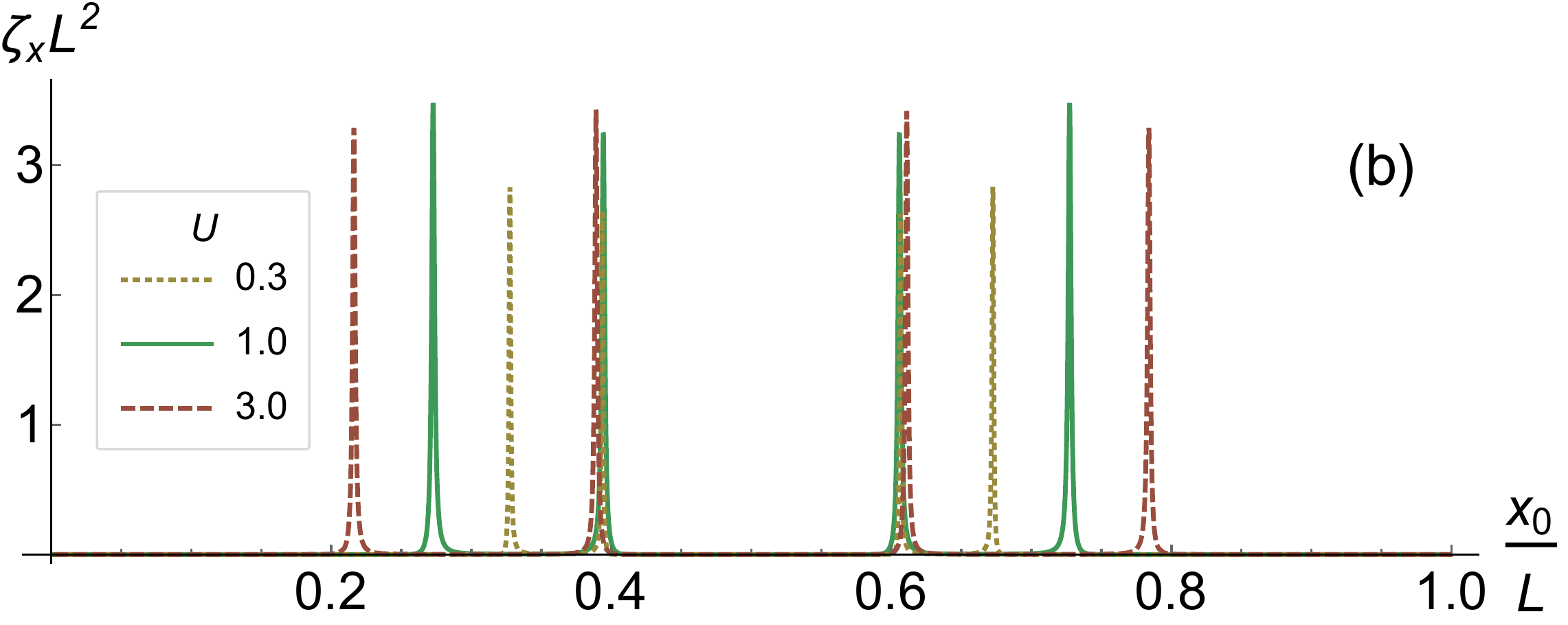}}
\end{tabular}
\caption{(Color online) The dependence of the $z$-component (a) and $x$-component (b) of the spin order parameter on the probe position for several values of the \textit{e-e} interaction amplitude $U$.}
\label{fig3}
\end{figure}

The most illustrative characteristic of the emergence and evolution of the spin density components is the spin order parameter defined as $\zeta_{\gamma}(x_0)=\int_0^{L}dx\,s_{\gamma}^2(x)$, $\gamma \in \{x,z\}$. Figure~\ref{fig3} shows its dependence on the probe position $x_0$ for three values of the \textit{e-e} interaction parameter $U$. The $x$-component has the form of four narrow peaks of width $\delta x_0$. Each peak corresponds to the consecutive transition of an electron from one side of the probe to the other. There are two bands of spin polarizations in the $z$ direction. The width of the spin polarization bands, as well as the magnitude of the polarization, depends dramatically on the \textit{e-e} interaction strength, vanishing as the interaction is reduced.

\begin{figure}
\includegraphics[width=0.9\linewidth]{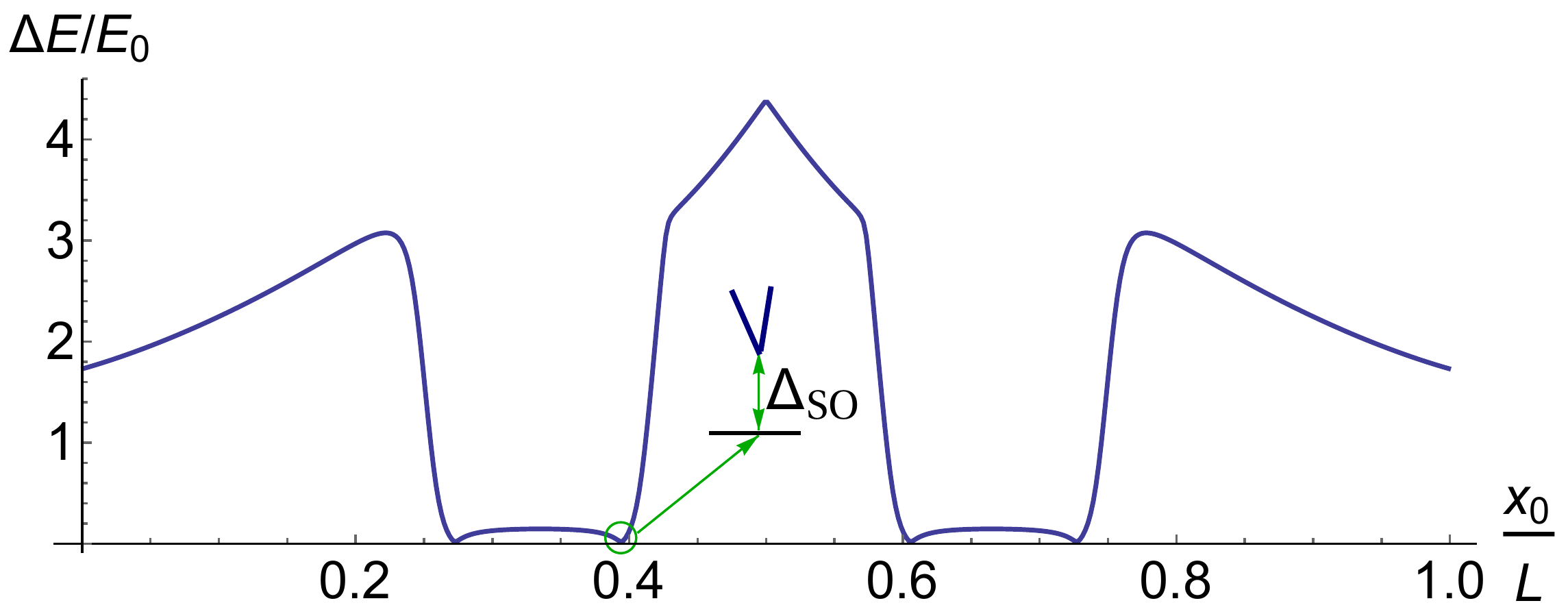}
\caption{(Color online) The energy gap between the ground state and the first excited level in the many-electron spectrum as a function of the probe position $x_0$. The inset shows the level splitting $\Delta_\mathrm{SO}$ at the avoided crossing of the terms.}
\label{fig4}
\end{figure}

The switching between the different spin states can be considered as an analogue of the STT. Figure~\ref{fig4} shows the dependence of the energy spacing between the ground state and the first excited state, which differ in their spin structure, on the probe position. Four critical positions $x_0$, at which the level spacing drops to $\Delta_\mathrm{SO}\approx 0.03$ meV, correspond to the four transitions of electrons from one well to another.

The value of $\Delta_\mathrm{SO}$ determines the energy gap by which the definite spin state is protected in the immediate vicinity of the STT, but in the center of the polarization band the polarized ground state is separated from the closest excited non-polarized state by the total of exchange and Zeeman energies, which in our system is by an order of magnitude larger than $\Delta_\mathrm{SO}$. The energy splitting $\Delta_{SO}$ depends on the \textit{e-e} interaction strength. The analysis of the \textit{e-e} interaction effect on this energy as well as on the spin density distribution is given in the Appendix. It is based on a simplified model justified for the conditions close to the STT point.

Above we have neglected the electron interaction with the nuclei. The spin manipulation mechanism considered here is robust against this interaction if the characteristic energies due to the fluctuations $\delta B_n$ of the Overhauser field and the difference in the Overhauser field $\Delta B_n$ between the polarized and unpolarized states are smaller than the energy difference between the polarized and unpolarized states. The lower bound of this value is determined by the energy gap protecting the ground state, which is presented in Fig.~\ref{fig4} and can be estimated in the center of the polarization bands. This condition is well satisfied for our system, assuming that the $\delta B_n, \Delta B_n \sim 10-10^2$~mT.\cite{chekhovich2013nuclear}

In summary, we have shown that the spin state of the 1D QD can be effectively controlled by the charged tip of the SPM in the presence of an external magnetic field. The analysis of the changes in the spatial distribution of the charge and spin density in a 1D QD containing four electrons has revealed the presence of two bands of the probe position where the QD acquires the net spin polarization in the direction of the magnetic field. Outside these bands the spin density is absent. In the narrow transition regions between the polarized and non-polarized states, the spin polarization directed along the QD arises.

The spin-polarized state occurs due to the joint action of the probe potential that divides the QD into two tunnel-coupled quantum wells and the strong Hubbard-like correlation of electrons in the narrowest well. Because of the Coulomb repulsion an electron having occupied the narrowest well with the spin parallel to the magnetic field blocks the electrons with the opposite spin in the other well.

The width of the polarization bands is determined the \textit{e-e} interaction strength and magnetic field. The spin structure in the transition regions and their width are determined by the Rashba SOI induced by the electric field of the charged probe.

\acknowledgments
This work was partially supported by Russian Foundation for Basic Research (project No~14-02-00237) and Russian Academy of Sciences.

\section*{Appendix}

In this section the effect of the \textit{e-e} interaction strength on the formation of the spin polarized states is studied under the conditions where the system is close to the transition point. We develop an analytic model which allows one to calculate the energy splitting $\Delta_{SO}$ and the spin density for the arbitrary strength of the \textit{e-e} interaction. The model is restricted by considering two electrons, but the conclusions are applicable to a system with a larger number of electrons since the most important electron correlations are caused by two electrons occupying higher energy levels, while other electrons are weakly disturbed by the correlations near the transition point. The results obtained within this simplified model qualitatively well agree with specific calculations using the exact diagonalization method, presented in the main text.

The single-particle Hamiltonian is $H_{1p}=-\frac12\partial_{x}^2+\phi_{\mathrm{pr}}(x)$, with the probe potential $\phi_{\mathrm{pr}}(x)$ localized around $x=0$. The open boundary conditions are imposed at $x=-L_1$ and $x=L_2$. We restrict the consideration to a two-level model taken as the two lowest-lying single-particle terms $\epsilon_1$ and $\epsilon_2$. The corresponding single-particle eigenstates $\phi_1(x)$ and $\phi_2(x)$ are localized on the different sides of the probe, which happens when the probe is close to the center of the system. The wavefunctions are orthonormal. The parameters are chosen so that the system is close to the STT.

In the absence of \textit{e-e} interaction the ground state is $\phi_1 (x_1)\phi_1(x_2)\chi_{S}$, with $\chi_{S}$ being the singlet spin wavefunction. Now let us find the two-particle wavefunctions of the interacting system. The trial wavefunction of the singlet state is
\begin{eqnarray}
 \Psi_{S}(x_1,x_2)=\left[a(\phi_1(x_1)\phi_2(x_2)+\phi_1(x_2)\phi_2(x_1))\nonumber\right.\\
 \left.+b\phi_1(x_1)\phi_1(x_2)\right]\chi_{S}\;,
\end{eqnarray}
where the coefficients $a$ and $b$ are defined by minimizing the variational energy of the full Hamiltonian $H=\sum H_{1p}+U(x_1-x_2)$. They are 
\begin{equation}
 a=-\frac{1}{\sqrt{2}}\sqrt{1-\frac{A}{\sqrt{A^2+2}}},\;b=\frac{1}{\sqrt{2}}\sqrt{1+\frac{A}{\sqrt{A^2+2}}}\;,
\end{equation}
where $A=(\epsilon_2-\epsilon_1+J_1-J_0)/J_2$. The interaction integrals $J_i$ are specified below. The largest one is
\begin{equation}
 J_0=\int_{-L_1}^{L_2}dx_1dx_2\,U(x_1-x_2)\rho_1(x_1)\rho_1(x_2)\;,
\end{equation}
which gives the ``on-site'' interaction independent of the density overlapping between the wells. Other interaction integrals
\begin{eqnarray}
 J_1=\int_{-L_1}^{L_2}dx_1dx_2\,U(x_1-x_2)\left(\rho_1(x_1)\rho_2(x_2)\right.\\ \nonumber
 +\left.\phi_1(x_1)\phi_2(x_1)\phi_1(x_2)\phi_2(x_2)\right)\;,
\end{eqnarray}
\begin{equation}
 J_2=2\int_{-L_1}^{L_2}dx_1dx_2\,U(x_1-x_2)\rho_1(x_1)\phi_1(x_2)\phi_2(x_2)
\end{equation}
and
\begin{eqnarray}
 J_3=\int_{-L_1}^{L_2}dx_1dx_2\,U(x_1-x_2)\left(\rho_1(x_1)\rho_2(x_2)\right. \\ \nonumber
 -\left.\phi_1(x_1)\phi_2(x_1)\phi_1(x_2)\phi_2(x_2)\right)
\end{eqnarray}
are as small as the overlapping is.

The ground state energy equals
\begin{eqnarray}
\label{gsen}
 \epsilon_{\mathrm{S}}=2\epsilon_1+J_0+\frac12\Big(\epsilon_2-\epsilon_1+J_1-J_0\nonumber\\
 -\sqrt{(\epsilon_2-\epsilon_1+J_1-J_0)^2+2J_2^2}\Big)\;.
\end{eqnarray}
The wavefunction of the triplet states has the form
\begin{equation}
 \Psi_{T}(x_1,x_2)=\frac{1}{\sqrt{2}}\left(\phi_1(x_1)\phi_2(x_2)-\phi_1(x_2)\phi_2(x_1)\right)\chi_{T}\;,
\end{equation}
where $\chi_T$ stands for spin functions $\left|T_0\right>=\left|S=1,S_z=0\right>$ and $\left|T_\pm\right>=\left|1,\pm 1\right>$. The triplet energy is 
\begin{equation}
\epsilon_{\mathrm{T}}=\epsilon_1+\epsilon_2+J_3\pm B\;. 
\label{tren}
\end{equation}
In the absence of SOI, the STT corresponds to the intersection of $\epsilon_{\mathrm{S}}$ and $\epsilon_{\mathrm{T}}$ terms. The system can be driven to the STT by the variation of the \textit{e-e} interaction amplitude $U$. Due to the different symmetry of the orbital wave functions, \textit{e-e} interaction contributes differently to the energy of singlet and triplet states, decreasing the ST energy levels spacing. A variation in $U$ could be realized in electrostatically formed quantum dots by squeezing the dot in the transverse direction by changing the potential of nearby gates. Indeed, the Coulomb interaction energy, estimated as $(2e^2/\varepsilon L)\ln|L/a|$, increases as the transverse length $a$ is reduced.

\begin{figure}
\includegraphics[width=0.9\linewidth]{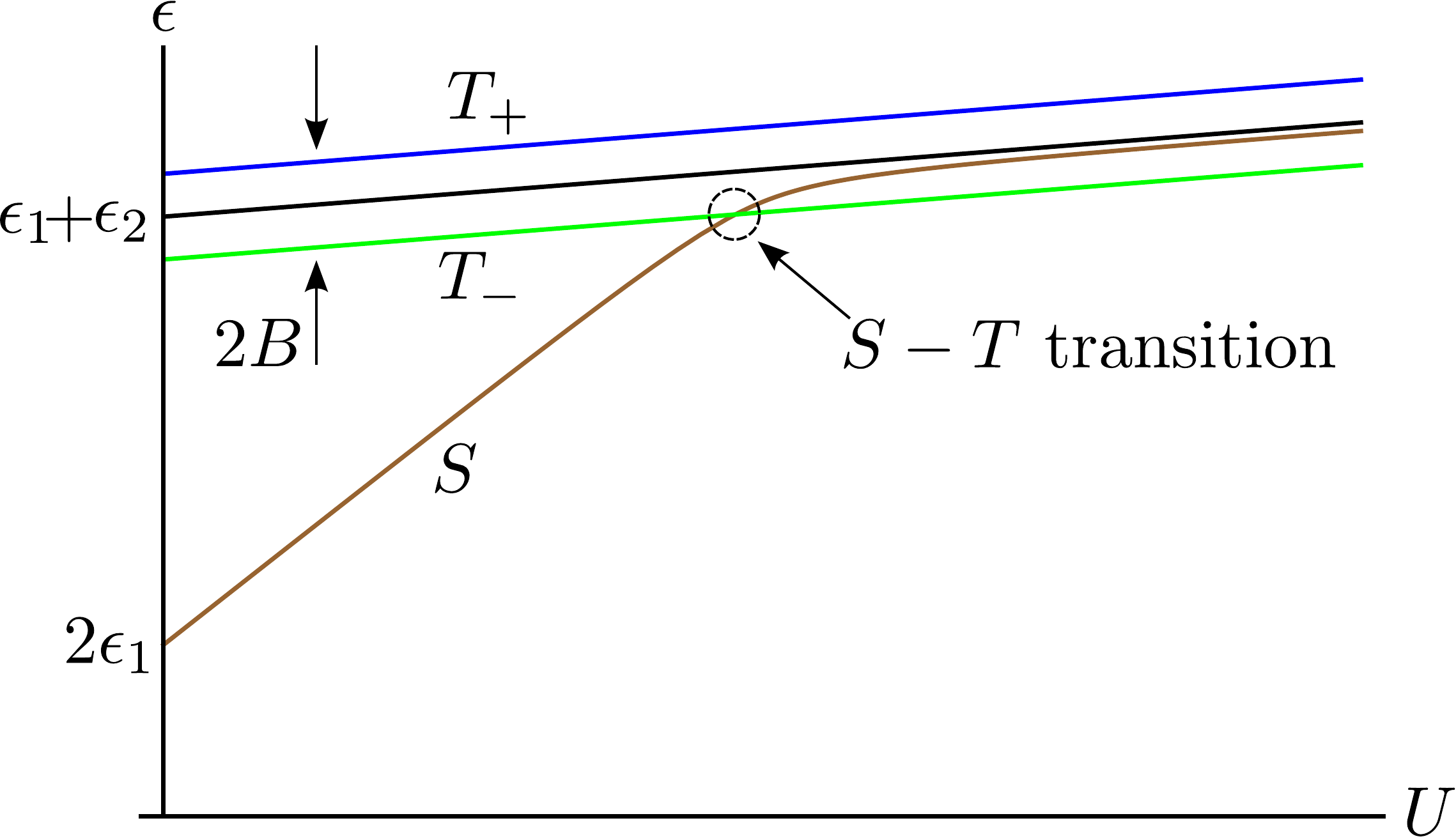}
\caption{The energy spectrum of our model system as a function of \textit{e-e} interaction strength.}
\label{sup_fig1}
\end{figure}

\begin{figure}
\includegraphics[width=0.9\linewidth]{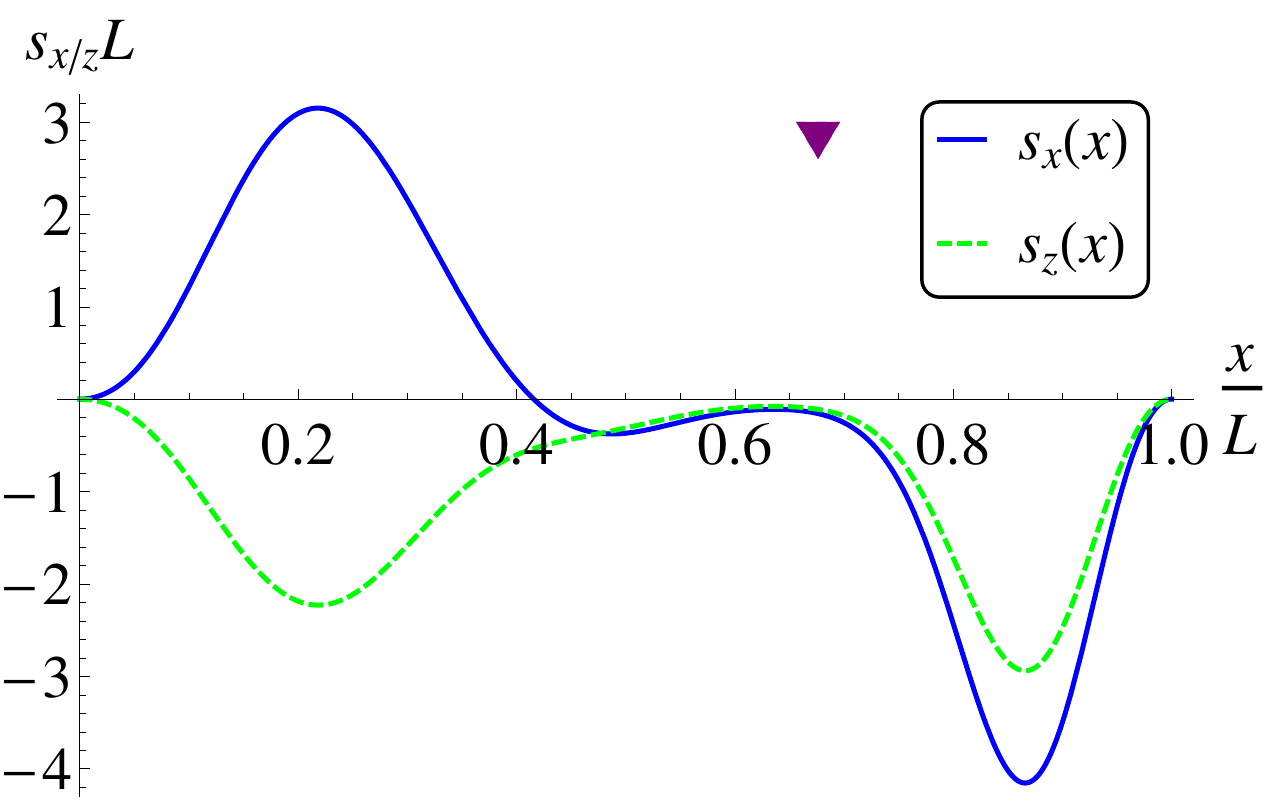}
\caption{The exact diagonalization results for the spin density components of the 2-electron 1D QD. The system parameters are as follows, the length $L=40 a_B$, the probe charge $Q=e$, the probe position is $x_0=0.676L$, the probe height above the QD is $z_0=4 a_B$, the magnetic field $B=3.6$ $\mu$T, which corresponds to $\hbar \omega_c=4 \cdot 10^{-5}E_0$, the longitudinal quantization energy $E_0=9.7$ $\mu$eV.}
\label{sup_fig2}
\end{figure}

The energy spectrum of our model as a function of $U$ is presented in Fig.~\ref{sup_fig1}. The singlet term first increases linearly as $U$ grows. This corresponds to the gradual occupation of the $\phi_2(x)$ state. As soon as this state is fully occupied, the transition of one electron from the left to the right side of the quantum wire is completed. The transition of a single electron from one well to another is the distinctive signature of the \textit{e-e} interaction, because in the non-interacting system electrons transit in pairs. The singlet term asymptotically approaches the $T_0$ term upon further increase of $U$ and intersects the $T_{-}$ term if the magnetic field exceeds some critical value $B_{c}$. The spin polarization along the $B$ direction appears, the transition to the polarized state being abrupt (a sharp step as a function of $U$). Eqs.~(\ref{gsen}) and~(\ref{tren}) define the critical value of the system parameters and the width of the polarization bands. The value of $B_{c}$ can be made rather small at sufficiently strong \textit{e-e} interaction (as is demonstrated by Fig.~\ref{sup_fig2}).

Now take into account the SOI of Eq.~(4) of the main text to the first order of its strength. The only non-zero SOI matrix elements are between $\left|S\right\rangle$ and $\left|T_\pm\right\rangle$ states, so the ground state represents a linear combination between the three. At $B=0$ the contributions of $\left|T_\pm\right\rangle$ to the spin density cancel each other, which is why there is no spin response in the absence of magnetic field (none is expected, since the Hamiltonian that respects the time-reversal symmetry yields zero average value of the spin density in the non-degenerate ground state with integer full spin). However, in order for STT to occur the finite magnetic field $B_{c}$ is applied that effectively lifts the state $\left|T_{+}\right\rangle$ up, leaving us with a two-level model, with the matrix element
\begin{eqnarray}
&&\left\langle T_{-}|H_{\mathrm{SO}}|S \right\rangle= \nonumber \\
&&\frac{\alpha b}{2}\int_{-L_1}^{L_2}dx E_{z}(x)\left(\phi_1(x)\phi_2'(x)-\phi_1'(x)\phi_2(x)\right)
\label{soimat}
\end{eqnarray}
The ST level spacing is
\begin{equation}
\Delta E_\mathrm{ST}=\sqrt{(\epsilon_{\mathrm{T_{-}}}-\epsilon_{\mathrm{S}})^2+4|\left\langle T_{-}|H_{\mathrm{SO}}|S \right\rangle|^2}\;.
\end{equation}
The ST level splitting $\Delta_\mathrm{SO}$ is defined as the minimum value of $\Delta E_\mathrm{ST}$. It is achieved at  $\epsilon_{\mathrm{T_{-}}}\approx\epsilon_{\mathrm{S}}$ and is equal to
\begin{equation}
\Delta_\mathrm{SO}=2|\left\langle T_{-}|H_{\mathrm{SO}}|S \right\rangle|\;.
\label{soisplit}
\end{equation}
This formula along with Eq.~(\ref{soimat}) shows that $\Delta_\mathrm{SO}$ depends on the probe position and the \textit{e-e} interaction strength, via the coefficient $b$. With increasing the \textit{e-e} interaction $b$ decreases, so at $b \to 0$ the ST splitting vanishes similar to the case of a uniform quantum wire.\cite{gindikin2011electron}

Let us find the spin density distribution to compare it with the exact diagonalization results. With SOI included the ground state becomes
\begin{equation}
 \Psi_{\mathrm{GS}}(x_1,x_2)=\frac{1}{\sqrt{c^2+1}}\left(\Psi_{\mathrm{S}}(x_1,x_2)+c\Psi_{\mathrm{T_{-}}}(x_1,x_2)\right)\;,
\end{equation}
where
\begin{equation}
 c=\frac{\epsilon_{\mathrm{T_{-}}}-\epsilon_{\mathrm{S}}-\Delta E_\mathrm{ST}}{\Delta_{\mathrm{SO}}}\;.
\end{equation}
The spin density components are as follows,
\begin{equation}
 s_x(x)=A_{x}\left(\rho_1(x)-\rho_2(x)\right)\;,
\end{equation}
and
\begin{equation}
 s_z(x)=-A_{z}\left(\rho_1(x)+\rho_2(x)\right)\;,
\end{equation}
where $\rho_i(x)=|\phi_i(x)|^2$, $i=\overline{1,2}$, is a single-particle density. The fact that the $z$-component of the spin density repeats the charge density profile $\rho_1(x)+\rho_2(x)$ is quite predictable for the fully polarized state. The anti-symmetric combination  $\rho_1(x)-\rho_2(x)$ obtained for $s_x(x)$ is less obvious but agrees well with the exact diagonalization results, presented in Fig.~\ref{sup_fig2}.

The amplitude $A_{x}=2ac/(c^2+1)$ as a function of $U$ has the form of the peak reaching its maximum value of $1/\sqrt{2}$ in the vicinity of the STT. The peak width is determined by $\Delta_{\mathrm{SO}}$ of Eqs.~(\ref{soimat}),~(\ref{soisplit}). The amplitude $A_{z}=c^2/(c^2+1)$ has a step-like structure as a function of $U$, growing from 0 to 1 in the STT vicinity, with the saturation of the magnitude marking the onset of the spin-$z$ polarization band. The form of the step is smoothed by the SOI. These results qualitatively agree with the exact diagonalization solution for 4 electrons.

\bibliography{paper}

\end{document}